\documentclass[a4paper,12pt]{article}

\usepackage{amsmath,amssymb}
\usepackage{graphicx}

\newcommand{\dd}{{\mathrm d}}
\newcommand{\ceff}{{\mathcal{E}}}
\newcommand{\Ndash}{\textendash}

\input epsf

\begin{document}

\title{Coalescence of particles by differential sedimentation}
\author{P. Horvai, S.V. Nazarenko, and T.H.M. Stein\footnote{Electronic address: t.stein@warwick.ac.uk}
\\ \normalsize University of Warwick, Mathematics Institute, CV4 7AL Coventry, United Kingdom}

\maketitle

\begin{abstract}

We consider a three dimensional system consisting of a large number
of small spherical particles, distributed in a range of sizes and
heights (with uniform distribution in the horizontal direction).
Particles move vertically at a size-dependent terminal velocity.  They
are either allowed to merge whenever they cross or there is a size
ratio criterion enforced to account for collision efficiency.
Such a system may be described, in mean field approximation, by the
Smoluchowski kinetic equation with a differential
sedimentation kernel.
We obtain self-similar steady-state and time-dependent solutions to
the kinetic equation, using methods
borrowed from weak turbulence theory.  Analytical results 
are compared with direct numerical simulations
(DNS) of moving and merging particles, and a good agreement is found.

%%   THIS IS MORE APPROPRIATE IN THE INTRODUCTION THAN IN THE ABSTRACT %%
%%
%% In the setup where small particles are
%% continuously injected and large ones removed uniformly in 3D space, we
%% obtain a constant flux Kolmogorov-Zakharov distribution, which appears
%% to be non-local in the free-merging model, and therefore only relevant
%% in the model with forced locality.  We obtain unsteady solutions which
%% describe gelation, i.e.\ creation of infinite particles in finite
%% time.  We find steady height-dependent solutions in which particles
%% are injected at some height and removed at a different height.

\end{abstract}

\section{Introduction}

We consider spherical particles in a viscous flow.  The particles move
vertically with their terminal velocity arising from the balance of
the gravitational effect (fall or buoyancy) and viscous drag.  
Since, in general, particles of different sizes
rise or fall with different velocities, their trajectories can cross and
merging can happen.  Realistic models of particle merging are
quite involved and in the present text we are going to consider only
two very simplified models: either any two particles whose trajectories
cross merge, which we shall refer to as ``free merging'',
or merging is restricted to particles of similar sizes (i.e.\ small
particles avoid big ones due to moving along flow streamlines bending
around the big particle), which we call ``forced locality'' (defined in
Sect.~\ref{sec:coleff}).

It will turn out that, although our problem is very simple to state,
it is very rich in features.  The simplified model can be realized by
considering a sedimenting kernel in the Smoluchowski coagulation
equation.  We will derive solutions to this equation analytically, and
we examine the validity of such solutions with direct numerical 
simulations (DNS), in which we let particles evolve individually according
to certain rules for collisions and we study their overall size distribution.  
We shall study different stationary regimes, either in time $t$ or 
in the vertical coordinate $z$, and we
will discuss self-similar solutions and study the role of local and
non-local merging. Whereas time dependent solutions of the sedimenting kernel have
received a lot of attention in the literature \cite{dong1,dong2,lee}, the study of height
dependence -- also treated here -- is more rare.

The process we discuss is usually referred to as differential
sedimentation and has been linked to experimental results \cite{hunt}
and is used to predict rain initiation time~\cite{pruppacher,falk}.
In particular, the model admits a power law distribution consistent
with experimental data for aerosols \cite{pruppacher}.  In our
discussion, we will obtain this power law as an exact result, rather
than by dimensional analysis used in previous discussions 
\cite{hunt,jeffrey}. We recognize this result as a Kolmogorov-Zakharov (KZ)
cascade of the volume integral, similar to the solutions that arise in
wave turbulence.  Solutions to the coagulation equation with a KZ
cascade have been studied in general \cite{cnzab,pushkin}, and with a
kernel describing galaxy mergers in particular \cite{vkon}.

We find that in the free-merging model the locality assumption
necessary in dimensional analysis and the KZ spectrum fail to hold
\cite{cnzab}. We will obtain an analytical solution for such
a non-local system, and verify this with DNS.  We will study
self-similarity for both the forced-locality model and the
free-merging model.  We will perform DNS for 
inhomogeneous solutions that are self-similar in the spatial
variable $z$.

The starting point of our analysis is to write a kinetic equation for
the coagulation process in Sect.~\ref{seccol}.  In
Sect.~\ref{sec:Kolmogorov} we find the Kolmogorov-Zakharov solution
for the kinetic equation. Sect.~\ref{sec:locality:sub:effective}
discusses the dominance of non-local interactions in the system. We
study self-similarity of our model in Sect.~\ref{sec:self-sim}, and we
analyze locality of such solutions in
Sect.~\ref{sec:locality-selfsim}, where we present numerical data.
Finally, we introduce a ``super-local'' model in
Sect.~\ref{sec:Burgers}, reducible to Burgers equation.

\section{The model}

Let us denote by $\sigma$ the volume of a spherical particle and by $r$ its
radius,
\begin{equation}
\label{sigma}
  \sigma
=
  \kappa r^3
\ ,
\qquad\qquad
	\kappa
=
	4\pi/3
\ .
%  r =
%  \left(\frac{3}{4\pi}\right)^{1/3} \sigma^{1/3} \ .
\end{equation}

\begin{figure}[ht]
\begin{center}
\includegraphics[width=.4\textwidth]{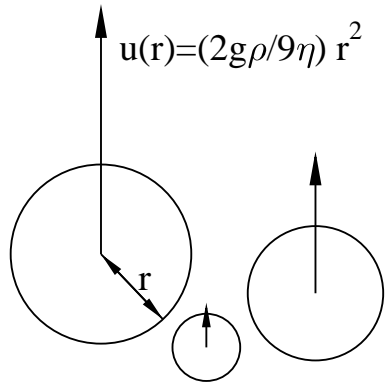}
\caption{A particle's terminal velocity $u$ is determined by its
radius $r$. Larger particles will have a larger terminal velocity,
depicted by the arrows, following definition \protect\eqref{u}.
(Created by T.H.M.Stein)}
\label{fig:termvol}
\end{center}
\end{figure}

The Stokes terminal velocity of a rigid sphere of radius $r$
with no slip boundary conditions
is given by the formula \cite{LandLif,pruppacher,clift}
\begin{equation}
  u(r)
=
  c r^2
\ ,
\qquad\qquad
	c
=
	\frac{2g(\rho_f-\rho_p)}{9\eta_f}
\ ,
\label{u}
\end{equation}
where $g$ is the free fall acceleration, $\rho_f$ and $\rho_p$ are the
density of the surrounding fluid and the particle respectively,
and $\eta_f$ is the dynamic viscosity of the surrounding fluid. 

Experimentally, the formulae \eqref{u} are valid  
for air bubbles in water at $20^{\circ}\mathrm{C}$ with $r < 1\mathrm{mm}$, 
and these bubbles can be considered spherical. Slip-flow corrections
can be necessary for other gases and fluids \cite{clift}.
The following data for water droplets and particles in the atmosphere can be found in Pruppacher and Klett~\cite{pruppacher}. For droplets, corrections
to \eqref{u} are necessary when $r > 30\mu\mathrm{m}$, 
which changes the formula's dependence on $r^2$. They can be considered 
spherical for radii up to $535\mu\mathrm{m}$.
For atmospheric particles, \eqref{u} can be considered
to depend on $r^2$ for large particles. However, atmospheric particles
are generally not spherical and will thus require 
other corrections.

Despite physical complications, we will assume \eqref{u} and 
\eqref{sigma}, and we will express both in terms of volume $\sigma$,
\begin{equation}
\label{u1}
  r(\sigma)
=
  \kappa^{-1/3} \sigma^{1/3}
\ ,
\qquad\qquad
  u(\sigma)
=
  c \kappa^{-2/3} \sigma^{2/3}
\ .
\end{equation}

We compute this model using 
direct numerical simulations in a periodic
box of $10\times10\times10\mathrm{cm}$ with particles that are defined
by their $x$-, $y$-, and $z$-coordinates and by their volume $\sigma$.
At each time step the particles move according to their fixed terminal 
velocity, using definition~\eqref{u}.  We fix our parameter $c$ 
such that a particle of radius $0.1\mathrm{cm}$
moves upwards with velocity $20\mathrm{cm} \mathrm{s}^{-1}$, 
which resembles the situation of air bubbles in water \cite{clift}. 

The particles are generated at a range of small
$\sigma$, with their smallest volume
$\sigma_0\approx4.2\cdot10^{-6}$cm$^3$, equivalent to a radius
$r=0.01\mathrm{cm}$. They are removed from the system once 
they become larger than $10^3\sigma_0$, or
$r\sim 1\mathrm{mm}$ and are assumed to be spherical at all sizes for
computational purposes. With different velocities, the particle
trajectories may cross, and depending on the rules of interaction they
can then merge. These rules are governed by collision efficiency,
which will be explained in Sect.~\ref{seccol}.

\subsection{The kinetic equation}
\label{seccol}

We suppose that the distribution of particles can be adequately
characterized by density $n(\sigma,z,t)$ (the number of
particles $N$ of volume between $\sigma$ and $\sigma+\dd\sigma$, per fluid 
volume $V$ per $\dd\sigma$, at the vertical coordinate $z$ and at instant $t$).
In particular we suppose here that the dependence of particle
distribution on the horizontal coordinates can be averaged out.  This
hypothesis is valid if the dynamics do not lead to strongly
intermittent distribution in the horizontal directions, for example if
the fluid is well mixed in the horizontal directions.
Our numerical simulations appear to support such a mean field approach well, 
and in future work it would be interesting to
examine theoretically why this is the case.

The goal of this section is to derive a kinetic equation for $n$ -- also
called Smoluchowski coagulation equation~\cite{smol} -- using a kernel 
describing differential sedimentation. We
write the collision integral, which expresses simply the fact that two
particles of volumes $\sigma_1$ and $\sigma_2$, with
$\sigma_1+\sigma_2=\sigma$, can merge to give a particle of volume
$\sigma$ (inflow), or a particle with volume $\sigma$ can merge with any
other particle of volume $\sigma_1 > 0$ and give a particle with volume
$\sigma_2=\sigma+\sigma_1 $ (outflow).  Also, we determine the
cross-section of interaction between two particles by the condition that
particles merge upon touching, that is if their centers are at a
distance at most $r_1+r_2$, which gives the geometric cross-section
of $\pi(r_1+r_2)^2$.  Finally the collision rate between particles of
volume $\sigma_1$ and $\sigma_2$ is taken to be proportional to their
relative velocities $|u(\sigma_1)-u(\sigma_2)|$ and to their number
densities $n_1$ and $n_2$, which is a mean field type hypothesis.

The left hand side of the kinetic equation contains the advection
term $\partial_t n + u \partial_z n$, which we shall also denote as
the total derivative $\dd n/\dd t$, while on the right hand side we
put the collision integral. Note also the shorthand
$n=n(\sigma,z,t)$, $u=u(\sigma)$, $n_1=n(\sigma_1,z,t)$,
$u_1=u(\sigma_1)$, $r_1=r(\sigma_1)$ and similar for $n_2$, $u_2$
and $r_2$. Thus we find
\begin{align}
\label{eq:kinold}
  \partial_t n + u \partial_z n &
= 
\\
\nonumber
  & +\frac{1}{2}\int_0^{\sigma}
  	\dd\sigma_1 \dd\sigma_2 \,
  |u_2-u_1| \pi (r_1+r_2)^2 n_1 n_2
         \delta(\sigma-\sigma_1-\sigma_2)
\\
\nonumber
	& -\frac{1}{2}\int_0^{+\infty}
  	\dd\sigma_1 \dd\sigma_2 |u-u_2| \pi (r+r_2)^2 nn_2
         	\delta(\sigma_1-\sigma-\sigma_2)
\\
\nonumber
	& -\frac{1}{2}\int_0^{+\infty}
   	\dd\sigma_1 \dd\sigma_2 |u-u_1| \pi (r+r_1)^2 nn_1
        	\delta(\sigma_2-\sigma-\sigma_1)
\ .
\end{align}
It is useful to express the $u$ and $r$ in terms of $\sigma$ 
using \eqref{u1},
\begin{flalign}
\label{eq:kinetic}
  \partial_t n + c \kappa^{-2/3} \sigma^{2/3} \partial_z n
=
\\
\nonumber
  \frac{c \kappa^{-4/3} \pi}{2} \int_0^{+\infty} \hspace{-3mm} \dd\sigma_1 
  \int_0^{+\infty} \hspace{-3mm} \dd\sigma_2 \hspace{3mm}
  & |\sigma_2^{2/3}-\sigma_1^{2/3}| (\sigma_1^{1/3}+\sigma_2^{1/3})^2 n_1 n_2
         \delta(\sigma-\sigma_1-\sigma_2)
\\
\nonumber
 	- & |\sigma^{2/3}-\sigma_2^{2/3}| (\sigma^{1/3}+\sigma_2^{1/3})^2 nn_2
         \delta(\sigma_1-\sigma-\sigma_2)
\\
\nonumber
  - & |\sigma^{2/3}-\sigma_1^{2/3}| (\sigma^{1/3}+\sigma_1^{1/3})^2 nn_1
         \delta(\sigma_2-\sigma-\sigma_1)
\ .
\end{flalign}
Let us introduce the interaction kernel $K(\sigma_1,\sigma_2)$, 
\begin{equation}
\label{eq:kernel}
  K(\sigma_1,\sigma_2)
=
  \frac{c \kappa^{-4/3} \pi}{2} |\sigma_2^{2/3}-\sigma_1^{2/3}|
  (\sigma_1^{1/3}+\sigma_2^{1/3})^2 
\ ,
\end{equation}
which for a general kernel $K$ reduces Eq.~\eqref{eq:kinold} to the
Smoluchowski equation.  It is useful to note that our
kernel~\eqref{eq:kernel} is homogeneous in $\sigma$, with
$K(\zeta\sigma_1,\zeta\sigma_2)=\zeta^{4/3}K(\sigma_1,\sigma_2)$.  We
also introduce the collision rates
\begin{equation}
\label{eq:rates}
  R_{\sigma 12}
=
  K(\sigma_1,\sigma_2) n_1 n_2 \delta(\sigma-\sigma_1-\sigma_2)
\end{equation}
with $R_{1\sigma 2}$, $R_{2\sigma 1}$ defined analogously.  Now the
RHS of Eq.~\eqref{eq:kinetic} can be written in a compact form
\begin{equation}
\label{eq:kinetic-sym}
  \frac{\dd n}{\dd t}
=
  \int_0^{+\infty} \hspace{-3mm} \dd\sigma_1
  \int_0^{+\infty} \hspace{-3mm} \dd\sigma_2\,
    (R_{\sigma 12} - R_{1\sigma 2} - R_{2\sigma 1})
\ .
\end{equation}

\subsection{Characteristic timescales}

We study the physical relevance of Eq.~\eqref{eq:kinetic} by comparing its
characteristic time $\tau_{ds}$ with the characteristic residence time
in a typical system, $\tau_g = L/u$, where $L$ is the vertical extent
of the system, and $u$ is as in Eq.~\eqref{u1}. To find $\tau_{ds}$,
we note that $n \sim \frac{N}{\sigma V}$ and we introduce the volume
fraction $v \sim \frac{N\sigma}{V}$, so that:
$$
	n
\sim
	\frac{v}{\sigma^2}
\ .
$$
Now, using the kinetic equation \eqref{eq:kinetic} we can write
\begin{equation}
\label{eq:timescale}
	\frac{1}{\tau_{ds}}
=
	c \kappa^{-4/3} \pi \sigma^{2+2/3+2/3-1} \frac{v}{\sigma^2}
=
	c \kappa^{-4/3} \pi \sigma^{1/3} v
\ .
\end{equation}
Thus we find the following relation between the characteristic times:
\begin{equation}
\label{eq:timerelation}
	\frac{\tau_g}{\tau_{ds}}
=
	\frac{L c \kappa^{-4/3} \pi \sigma^{1/3} v}{c \kappa^{-2/3} \sigma^{2/3}}
\approx
	\frac{2L}{r} v
\ ,
\end{equation}
where we recall that $\sigma^{1/3}=\kappa^{-1/3} r$ and approximate $\kappa^{-1/3} \pi \approx 2$. From \cite{pruppacher} we find that for a cumulus cloud, typically $L \sim 10^3 \mathrm{m}$, $r\sim 10^{-5} \mathrm{m}$, and $v\sim 10^{-6}$. Thus, we find that $\tau_g/\tau_{ds} \sim 10^2$, which implies that the kinetic equation is relevant in a cloud system with gravity when we regard time and length scales.

\subsection{Collision efficiency}
\label{sec:coleff}

The kinetic equation~\eqref{eq:kinetic} allows merging of particles of
any sizes, without any discrimination. We shall refer to this case as
``free merging''.  More realistically one should
also take into account the collision efficiency between particles. We
define collision efficiency $\ceff_{12} = \ceff(\sigma_1,\sigma_2)$
between particles of volumes $\sigma_1$ and $\sigma_2$ as a number
between 0 and 1, which enters the collision integral by multiplication 
with the collision rates $R$, so $R_{\sigma 12}$ would be replaced by
$R_{\sigma 12} \ceff_{12}$ and more generally for example the
integrand of Eq.~\eqref{eq:kinetic-sym} would become $R_{\sigma 12}
\ceff_{12} - R_{1\sigma 2} \ceff_{\sigma 2} - R_{2\sigma 1}
\ceff_{\sigma 1}$.

\begin{figure}[ht]
\begin{center}
\includegraphics[width=.4\textwidth]{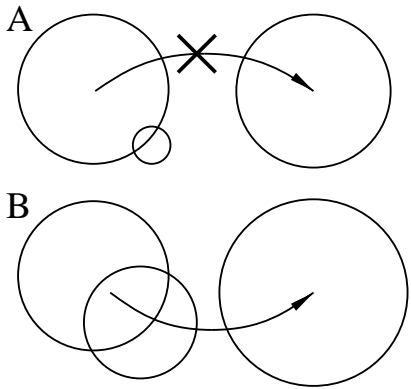}
\caption{Without applying the efficiency kernel $\ceff$, particles merge
whenever they cross. Including $\ceff$ with small $q$, only situation
B is allowed, i.e.\ only particles of similar size may merge; particles of
dissimilar size (situation A) are allowed to cross one another without
merging. (Created by T.H.M.Stein)} \label{fig:colleff}
\end{center}
\end{figure}

In particular, one could restrict merging to particles of similar
sizes, taking into account that small particles cannot collide with
much larger ones because they bend around them along the fluid
streamlines. In the simplest such model which will be considered
later in this paper,
\begin{equation}
\label{def:local-efcncy}
  \ceff_{12}
=
  \begin{cases}
    1 & \text{if $1/q < \sigma_1/\sigma_2 <q$,}\\
    0 & \text{otherwise,}
  \end{cases}
\end{equation}
where $q>1$ is the number representing the maximal volume ratio for
the particle merging. Compared to a more involved form of collision
efficiency used by Valioulis et al.~\cite{VLP}, the simplified kernel 
we use mimics the behavior for particles with $r=0.01\mathrm{cm}$  
which is similar to the regime we study numerically. We will refer 
to the model with finite $q$ as ``forced locality''.

\subsection{Scaling argument}

For our simple setup one could derive a steady state solution 
merely by physical and dimensional arguments, following Friedlander~\cite{fried2}, Jeffrey~\cite{jeffrey}, and Hunt~\cite{hunt}.
The main remark is that at steady state, the system has a constant flux of volume.  The total volume of particles per unit volume of fluid that passes from particles smaller than $\sigma$ to particles greater than $\sigma$ is of the order:
\begin{equation}
\label{eq:volume-int}
	\int_\sigma^{2\sigma} \frac{\dd n}{\dd t} s \dd s 
\ .
\end{equation}
We can estimate from the kinetic equation~\eqref{eq:kinetic-sym} and equations~\eqref{eq:rates} and ~\eqref{eq:kernel} that $\dd n / \dd t \sim \sigma^2 R$, with $R \sim K n^2 \sigma^{-1}$ and $K \sim \sigma^{4/3}$. If we assume that $n \sim \sigma^{\nu}$, we find that $\dd n / \dd t \sim \sigma^{7/3 + 2 \nu}$, and we obtain the scaling $\sigma^{13/3+2\nu}$ for the volume flux~\eqref{eq:volume-int}. For constant flux, we arrive at $\nu = -13/6$, or $n \sim \sigma^{-13/6}$. 
Naturally, the dimensional analysis assumes locality of interactions.

%% We need to bear in mind that this is effectively a solution only if
%% our initial hypothesis of local interaction, needed to derive this
%% solution, is not violated by our solution.  Questions of locality are
%% discussed in full detail in Sect.~\ref{sec:locality}.

\section{Kolmogorov-Zakharov solution}
\label{sec:Kolmogorov}

One of the simplest questions one can ask with respect to the kinetic
equation~\eqref{eq:kinetic} is if it allows for a scaling stationary
solution of non-zero flux.  Such a solution, if one exists, is called
a Kolmogorov-Zakharov (KZ)  spectrum because, like in the classical
Kolmogorov spectrum, it corresponds to a cascade of a conserved
quantity (total volume occupied by particles in our case)~\cite{cnzab,vkon}.  
In this section we investigate the scaling exponent 
and existence of such solutions.

\subsection{Zakharov transform}

A derivation of the KZ solution can be achieved through the
technique of the Zakharov transform~\cite{cnzab,ZLF}.  Let us
consider a steady state (i.e.\ time and space independent) solution of
Eq.~\eqref{eq:kinetic} of form $n \sim \sigma^\nu$, and let us aim to
find $\nu$.  Note that this is a reasonable thing to look for, since
we can easily see from Eq.~\eqref{eq:kinetic} that our collision integral
is a homogeneous function in the $\sigma$ and in the $n$.

We start by expanding our collision rates from equation~\eqref{eq:rates} using equation
\eqref{eq:kernel}, and obtain the following equation in $\sigma$:
$$
  R_{\sigma 12}
=
  \frac{c \kappa^{-4/3} \pi}{2} |\sigma_2^{2/3}-\sigma_1^{2/3}|
  (\sigma_1^{1/3}+\sigma_2^{1/3})^2
  \sigma_1^\nu \sigma_2^\nu \delta(\sigma-\sigma_1-\sigma_2)
$$
where $R_{1\sigma 2}$ and $R_{2\sigma 1}$ are expanded similarly. We then continue by non-dimensionalising the rates $R$ by writing $\sigma_1$ as $\sigma'_1 \sigma$ and
$\sigma_2$ as $\sigma'_2 \sigma$, so
\begin{align}
\label{eq:R_prime}
  R_{\sigma 12} &
=
\\
\nonumber
  & \frac{c \kappa^{-4/3} \pi}{2} \sigma^{1/3 + 2\nu}
  |{\sigma'_2}^{2/3}-{\sigma'_1}^{2/3}|
  ({\sigma'_1}^{1/3}+{\sigma'_2}^{1/3})^2
  {\sigma'_1}^\nu {\sigma'_2}^\nu \delta(1-\sigma'_1-\sigma'_2)
\end{align}
and $R_{1\sigma 2}$ and $R_{2\sigma 1}$ are transformed in a similar way.

The Zakharov transform consists in passing in $R_{1\sigma 2}$ to
new variables $\tilde\sigma_1$ and $\tilde\sigma_2$ defined by
\begin{equation}
\nonumber
  \sigma'_1 = \frac{1}{\tilde\sigma_1}
\ ,
\qquad\qquad
   \sigma'_2 = \frac{\tilde\sigma_2}{\tilde\sigma_1}
\ .
\end{equation}
This way, we obtain
\begin{align}
\label{R_tilde}
  & R_{1 \sigma 2}
=
\\
\nonumber
  & \frac{c \kappa^{-4/3} \pi}{2} \sigma^{2\nu+1/3}
  {\tilde\sigma_1}^{-1/3 - 2\nu}
  |{\tilde\sigma_2}^{2/3}-{\tilde\sigma_1}^{2/3}|
  ({\tilde\sigma_1}^{1/3}+{\tilde\sigma_2}^{1/3})^2
  {\tilde\sigma_2}^\nu {\tilde\sigma_1}^\nu \delta(1-\tilde\sigma_1-\tilde\sigma_2)
\ .
\end{align}
A similar expression is derived for $R_{2\sigma 1}$. 

Combining the transformed terms and dropping primes and tildes, we transform the compact kinetic equation~\eqref{eq:kinetic-sym}
\begin{equation}
\nonumber
  0
=
  \int_0^{+\infty} \hspace{-3mm} \dd\sigma_1
  \int_0^{+\infty} \hspace{-3mm} \dd\sigma_2 \
    (1-\sigma_1^{-10/3-2\nu}-\sigma_2^{-10/3-2\nu})
    R_{\sigma 12}
\ .
\end{equation}
Here, we note that the integration variables for $R_{1 \sigma 2}$ become $\dd\sigma_1 \dd\sigma_2 = \sigma^2 {\tilde\sigma_1}^{-3} \dd \tilde\sigma_1 \dd \tilde\sigma_2$, with a similar transformation in $R_{2\sigma 1}$.
Now, if we choose $\nu$ such that $-10/3-2\nu=1$, then we have the
factor $\delta(1-\sigma_1-\sigma_2) (1-\sigma_1-\sigma_2)=0$ appearing
in the integrand, which solves the equation, i.e. $\nu =
-13/6$ is the candidate for the KZ exponent. This method of derivation can be applied
to various kernels for the Smoluchowski equation~\cite{cnzab}.

Let us  note that our exponent $\nu$ is that of $n(\sigma)$.
In literature, one commonly finds the radius
distributions,  $n(r)$,  which can be expressed in terms of  $n(\sigma)$
from the relationship $n(\sigma)\dd\sigma = n(r)\dd r$.
Thus,  $n(r) = n(\sigma) \dd\sigma/\dd r \propto r^{3\nu} r^2 =
r^{3\nu+2}$, and therefore $\nu_r = 3\nu+2 = -9/2$~\cite{jeffrey}.

However, the KZ spectrum is only a true solution of
Eq.~\eqref{eq:kinetic} if the collision integral on the RHS of this
equation (prior to the Zakharov transformation) converges. This
property is called locality, and it physically means that the particle
kinetics are dominated by mergings of particles with comparable (rather
than very different) sizes. Convergence of the collision integral on
general power-law distributions will be studied in
Appendix~\ref{sec:locality}.  We will see that (without modifying the
model to enforce locality) the $-13/6$ scaling exponent gives rise to
non-local interaction between the particles both with the smallest and
the largest particles and, therefore, the KZ spectrum is not a valid
solution in this case.

\begin{figure}[t]
\begin{center}
\includegraphics[width=.75\textwidth]{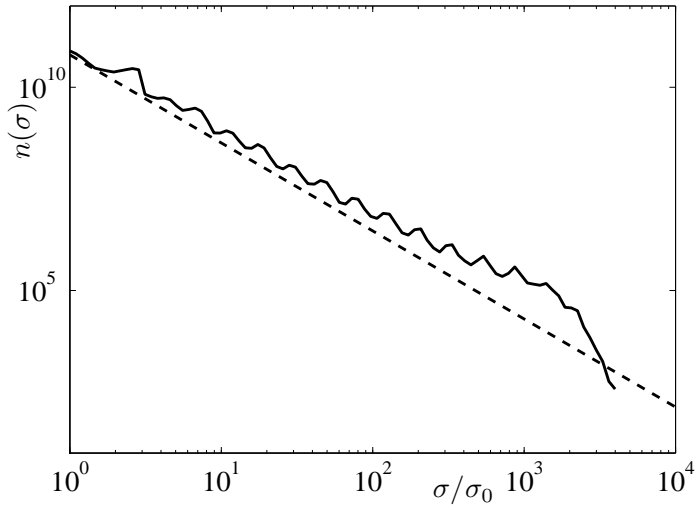} % [height=3in,width=4in]
\caption{Distribution of particle volumes averaged over several times after
140,000 time steps for the forced locality situation with $q=2$. 
The dashed slope represents the $-13/6$ KZ spectrum (compare with \protect\cite{VLP}).} 
\label{fig:kolzakh}
\end{center}
\end{figure}

\subsection{KZ spectrum in the system with forced locality}

Locality of interactions, and therefore validity of the KZ solution,
are immediately restored if one modifies the model by introducing the
local collision efficiency kernel as in definition~\eqref{def:local-efcncy}.
This kernel is a homogeneous function of degree zero in $\sigma$ and,
therefore, the KZ exponent obtained via the Zakharov transformation
remains the same. In Fig.~\ref{fig:kolzakh} we can see that the
Kolmogorov-Zakharov scaling appears in a system with forced
locality.

\section{Kinetics dominated by non-local interactions}
\label{sec:locality:sub:effective}

As an alternative, we may assume that the dominant
interactions are non-local and find a cut-off dependent stationary
solution. This is relevant if it is not desirable to use the collision
efficiency models which guarantee locality (for instance using 
the kernel~\eqref{def:local-efcncy}). In this case one should accept the fact 
the kinetics are dominated by non-local interactions, 
and that the low-$\sigma$ or/and high-$\sigma$ cut-offs dominate 
the collision integral.  In fact, such a
non-locality can allow us to significantly simplify the kinetic
equation and reduce it to a differential equation form. As shown in
Appendix~\ref{sec:locality}, contribution to the collision integral
from non-local interactions with the smallest particles ($\sigma_1 \ll
\sigma$) is
\begin{equation}
\label{def:c1}
  -c_1 \partial_\sigma(\sigma^{4/3} n)
\ ,
\qquad\text{where}\qquad
  c_1
=
  \int_{\sigma_{\min}} n_1 \sigma_1 \dd\sigma_1
\ .
\end{equation}
where we have dropped the explicit dependence of the upper integration
limit on $\sigma$, since the integral is divergent as ${\sigma_{\min}}
\to 0$ (this is the hypothesis of non-locality), so the dependence on
the upper bound is a sub-dominant contribution.

The contribution to the collision integral from non-local
interactions with the largest particles ($\sigma_1 \gg \sigma$) is
\begin{equation}
\label{def:c2}
  -c_2 n
\ ,
\qquad\text{where}\qquad
  c_2
=
  \int^{\sigma_{\max}} n_1 \sigma_1^{4/3} \dd\sigma_1
\ .
\end{equation}
Similarly to above, here the lower integration bound is omitted.

%If at one end interaction is non-local but on the other side it is
%local, then for that side one deduces simply from the kinetic equation
%the contribution $$  \propto \pm \sigma^{7/3} n^2 $$
%where positive sign is for lower end and negative sing is for high
%end (clearly, interaction with large particles depletes particles of size
%$\sigma$, so for a steady state to exist the smaller particles must feed
%them ... out of steady state the sign for the lower contribution is
%not clear, perhaps undetermined, needs more thinking).

%If interaction at both ends is local then we need to take more care,
%since the crude approximation above is insufficient.  Instead we look
%for a first order differential operator which has the same homogeneity
%as the kinetic equation, that is $\sigma^{7/3} n^2 $, and which
%conserves the total volume $\int n \sigma \dd\sigma$, so that it must
%start with $\sigma^{-1} \partial_\sigma$.  This then imposes the
%operator
%\begin{equation}
%\label{eq:2-local}
%  -\sigma^{-1} \partial_\sigma (\sigma^{13/3} n^2)
%\end{equation}
%up to a constant prefactor.  The sign is there because the volume flux
%is always positive, since particles only merge.

%\subsection{What next ?}

%Locality can be forced through mechanism of inefficient collisions
%between particles of very disproportionate sizes, for example small
%particle flowin around big one ...

%\section{Stationary state with non-local interactions}
\label{sec:NL-stat}

%Here we do not suppose forced locality of interaction, so that
%interaction may be non-local either with the smallest or largest
%particles or with both, and we look for stationary states of the kinetic
%equation~\eqref{eq:kinetic}.  We shall only treat in detail the case
%of both ends non-local, since all other cases lead to not
%self-consistent solutions, as can be easily checked.  For one, the
%hypothesis of local interaction at both ends has been explored in
%Sect.~\ref{sec:Kolmogorov} and in view of
%Sect.~\ref{sec:locality:sub:scaling} it is found not to be
%self-consistent.

Putting these two formulae together, we obtain the following
effective kinetic equation for the cases when the non-local
interactions are dominant,
\begin{equation}
\label{eq:NL-effkin}
  \frac{\dd n}{\dd t}
=
  - c_1 \partial_\sigma(\sigma^{4/3} n) - c_2 n
\ ,
\end{equation}
where constants $c_1$, $c_2$ are defined in the formulae~\eqref{def:c1} and \eqref{def:c2}. 
Note that this equation~\eqref{eq:NL-effkin} is valid when the non-local
interactions with the smallest and with the largest particles give
similar contributions, as well as in cases when one type of
non-locality is dominant over the other.

In steady state $\dd n/\dd t = 0$ and the solution of the resulting
ordinary differential equation is
\begin{equation}
\label{eq:n-NLNL}
  n
=
  C \sigma^{-4/3} e^{\frac{3c_2}{c_1} \sigma^{-1/3}}
\ ,
\end{equation}
with $C$ being an arbitrary positive constant.  Note that the
constants $C$ and $c_2/c_1$ appearing in the
solution~\eqref{eq:n-NLNL} can be related to the ``physical'' data of
$\sigma_{\min}$, $\sigma_{\max}$ and $n(\sigma_{\min})$, through
Eqs.~\eqref{def:c1}, \eqref{def:c2} and \eqref{eq:n-NLNL}.  We obtain
\begin{equation}
\label{eq:2-NL_stat-sol}
  n(\sigma)
=
  n(\sigma_{\min})
  \frac{
    \exp\left[
      \left(\frac{\sigma}{\sigma_{\min}}\right)^{-1/3}
      \log\frac{\sigma_{\max}}{\sigma_{\min}}
    \right]}
    {\left( \frac{\sigma}{\sigma_{\min}}\right)^{4/3}
      \frac{\sigma_{\max}}{\sigma_{\min}}
    }
\ .
\end{equation}

The solution~\eqref{eq:n-NLNL} is interesting since it is not a pure
power law.
%  In particular we need to decide if it is self-consistent, that is if it
%really gives rise to non-locality at small and at large scales.
For large $\sigma$ we have $n \sim C \sigma^{-4/3}$ which is a limit
when absorption of the smallest particles is much more important than
being absorbed by the large particles, i.e.\ when the first term on the
LHS of Eq.~\eqref{eq:n-NLNL} is much greater than the second one.
This limit corresponds to a cascade of the number of particles (not their
volume!) which is a conserved quantity in this regime.

\begin{figure}[t]
\begin{center}
\includegraphics[width=.75\textwidth]{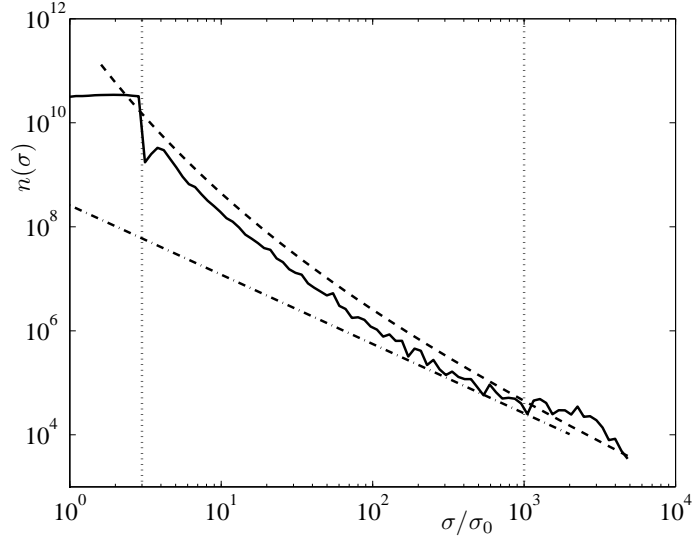} % [height=3in,width=4in]
\caption{Averaged distribution of particle sizes for the situation
without forced locality (``$q=\infty$'') after 200,000 time steps.  The
vertical dotted lines bound the inertial range at
$\sigma_{\min}=3\sigma_0$ and $\sigma_{\max}=10^3\sigma_0$. The
dashed curve represents the fit conform
eq.~\eqref{eq:2-NL_stat-sol}, with $\sigma_{\min}$ and $\sigma_{\max}$
given by the bounds of the inertial range, and
$n(\sigma_{\min})=1.5 \cdot 10^{10}$; the dash-dot slope represents a power
law of $\sigma^{-4/3}$.} \label{fig:nonloc}
\end{center}
\end{figure}

In Fig.~\ref{fig:nonloc} we show our numerical results for the
non-local model.  Particles are produced uniformly in space with volumes
ranging from $\sigma_0$ to $3\sigma_0$, and particle density within this
size range is kept constant in time.  Particles are removed from the system
once they reach $\sigma_{\max}=10^3\sigma_0$, with probability
$p(\sigma)=1-\exp^{-a(\sigma-\sigma_{\max})^4}$ with $a \ll 1$.
The original results have been averaged over neighbouring data
points to obtain the continuous graph in Fig.~\ref{fig:nonloc}. We
also used Eq.~\eqref{eq:2-NL_stat-sol} and find that with appropriate
parameters this solution fits the numerical data.

We can check our hypothesis of dominance of non-local
interactionsdirectly by counting the number of collisions within a certain
timeframe at statistical steady state. Namely, for each size bin we
count the number of collisions leading to a particle entering
the bin, and the number of collisions leading to a particle leaving 
the bin. We distinguish between local and non-local collisions using
the particle size ratio $q^*$, i.e. if $1/10 < q^* < 10$ we consider
the collision local, and non-local otherwise. For non-local collisions, 
we distinguish between a collision with a very large particle and a
very small particle. In the kinetic equation~\eqref{eq:kinetic} 
(which we do not rely on in our procedure) 
this would correspond to splitting the collision
integral as follows:
\begin{flalign}
\nonumber
  \frac{\dd n}{\dd t}
=
&+\int_{\sigma_{\min}}^{\sigma/q} \dd \sigma_1 f(\sigma_1,\sigma-\sigma_1)
 -\int_{\sigma_{\min}}^{\sigma/q} \dd \sigma_1 f(\sigma_1,\sigma)
\\
\label{eq:collcount}
&+\int_{\sigma/q}^{\sigma/2} \dd \sigma_1 f(\sigma_1,\sigma-\sigma_1)
 -\int_{\sigma/q}^{q\sigma} \dd \sigma_1 f(\sigma_1,\sigma)
\\
\nonumber
& - \int_{q\sigma}^{\sigma_{\max}} \dd \sigma_1 f(\sigma_1,\sigma)
\end{flalign}
where
$$
	f(\sigma_1,\sigma_2)
=
	K(\sigma_1,\sigma_2) n_1 n_2
\ .
$$

\begin{figure}[t]
\begin{center}
\includegraphics[width=.75\textwidth]{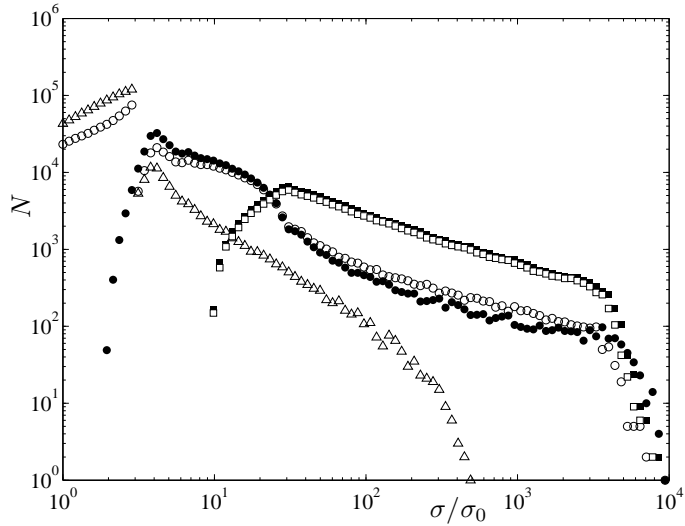} % [height=3in,width=4in]
\caption{Number of collisions $N$ per bin $[1.1^k \sigma_0, 1.1^{k+1} \sigma_0]$ over
10,000 time steps, which lead to a particle entering or leaving the bin. 
Triangles: contribution due to collisions with large particles; 
circles: contribution due to collisions with similar sized particles; squares: 
contribution due to collisions with small particles. 
Filled and open symbols correspond to number of particles entering and leaving
the bin respectively.}
\label{fig:collcount}
\end{center}
\end{figure}

We perform DNS and for each collision that occurs we count its
contribution to the different collision regimes as mentioned above. 
Our results are shown in Fig.~\ref{fig:collcount}. 
We notice that once collisions with small particles 
are counted at $\sigma/\sigma_0=q$, with $q=10$ in this figure, 
their contribution dominates almost immediately, and remains dominant for the
entire inertial domain. We can also see that collisions with larger particles
are only dominant in the forcing range $\sigma<3\sigma_0$, and collisions 
with similar sized particles only marginally dominates in the intermediate regime
for $3\sigma_0<\sigma<30\sigma_0$.

% which effectively
%gives rise to non-local interaction.  On the other hand for $\sigma$
%small $n$ grows as a stretched exponential, that is very fast, much
%faster than any power.  This also leads to non-local interaction.
%
%The constants $A$ and $c_2/c_1$ appearing in the solution
%\eqref{eq:n-NLNL} can be related to the ``physical'' data of
%$\sigma_{\min}$, $\sigma_{\max}$ and $n(\sigma_{\min})$, through
%Eqs.~\eqref{def:c1}, \eqref{def:c2} and \eqref{eq:n-NLNL}.  We obtain
%\begin{equation}
%\label{eq:2-NL_stat-sol}
%  n(\sigma)
%=
%  n(\sigma_{\min})
%  \frac{
%    \exp\left[
%      \left(\frac{\sigma}{\sigma_{\min}}\right)^{-1/3}
%      \log\frac{\sigma_{\max}}{\sigma_{\min}}
%    \right]}
%    {\left( \frac{\sigma}{\sigma_{\min}}\right)^{4/3}
%      \frac{\sigma_{\max}}{\sigma_{\min}}
%    }
%\ .
%\end{equation}

\section{Self-similar solutions}
\label{sec:self-sim}

KZ solutions studied in Sect.~\ref{sec:Kolmogorov} are valid
stationary solutions of the kinetic equation~\eqref{eq:kinetic} in the
systems modified by introduction of a local collision efficiency
(e.g. using the model~\eqref{def:local-efcncy}).  We have argued in
Sect.~\ref{sec:locality:sub:effective} that without such an enforced
locality the non-local interactions are dominant which results in a
prediction for the steady state given in Eq.~\eqref{eq:n-NLNL} and
which is qualitatively confirmed in direct numerical simulations of
the dynamics of particles.

However, both of these approaches assume homogeneity in space as well
as a sink at large volumes (i.e.\ removing particles from the system
when they reach a certain large size). These two conditions cannot be
made realistically consistent because there is not a physical
mechanism that could remove large particles from the bulk of the fluid.

Thus, it is more realistic to consider one of the following
solutions:
\begin{itemize}
\item time-dependent, height-independent solutions without a sink
\item height-dependent, time-independent solutions
  with a sink at a given height (i.e.\ 
  for bubbles in water an interface
  with air at a given maximum value of $z$).
\end{itemize}
Both situations can be described by self-similar solutions of
the kinetic equation~\eqref{eq:kinetic}.
In the following derivations of the self-similar solutions we will suppose locality,
in the sense that the dimensional analysis leading to the results
supposes no dependence on the cut-off scales $\sigma_{\min}$ and
$\sigma_{\max}$.  Validity of the locality hypothesis will have to be
examined {\em a posteriori}.

We will start by considering the particle model without forced
locality, and later we will proceed by adding the effect of local
collision efficiency followed by a super-local model leading
to Burgers equation.

%A crucial hypothesis is that of local interaction in the collision
%integral.  It is to be pointed out that simple considerations of
%dimensional analysis would fail if the integration bounds
%$\sigma_{\min}$ and $\sigma_{\max}$ had to be taken into consideration
%also.

\subsection{Height dependent solutions}
\label{sec:self-sim:sub:t-indep}

Let us start with the analysis of the time-independent state.  We look
for a solution $n$ that is self-similar in the sense that it verifies
the scaling relation
\begin{equation}
\label{eq:self-sim_t-indep}
  n(\sigma,z)
=
  z^\alpha h(z^\beta \sigma)
\ .
\end{equation}
To determine the exponents $\alpha$ and $\beta$ we need two
relationships.  The first one is that Eq.~\eqref{eq:kinetic} should
give an equation on $h$ as follows: introduce the self-similar variable
$\tau = z^\beta \sigma$ to replace all occurrences of $\sigma$, 
then Eq.~\eqref{eq:kinetic} can be written as
\begin{equation}
\label{eq:subs-tau}
  \tau^{2/3} z^{\alpha-\frac{2}{3}\beta-1}
  [\alpha h(\tau) + \beta \tau h'(\tau)]
=
  z^{2\alpha-\frac{7}{3}\beta} 
  \int_0^{+\infty} \hspace{-.3cm} \dd\tau_1 
  \int_0^{+\infty} \hspace{-.3cm} \dd\tau_2 \, (T_{\tau12}-T_{1\tau2}-T_{2\tau1})
\end{equation}
with the rate 
$$
	T_{\tau12} 
= 
\frac{c \kappa^{-4/3} \pi}{2} |\tau_2^{2/3}-\tau_1^{2/3}|
(\tau_1^{1/3}+\tau_2^{1/3})^2 h(\tau_1) h(\tau_2)
\delta(\tau-\tau_1-\tau_2)
$$ 
with $T_{1\tau2}$ and $T_{2\tau1}$ defined accordingly. 
We need to have equal powers of $z$ on both sides, which gives
$$
  \alpha - \frac{2}{3}\beta - 1
=
  2\alpha - \frac{7}{3}\beta
\ .
$$

The other relationship expresses constant flux of mass through a given
height $z$.  Since droplets of volume $\sigma$ move with speed
$u=u(\sigma)$, this flux is $\int n(z,\sigma) u \sigma \dd\sigma$.
With $h$ and $\tau$ this becomes $\int z^\alpha h(\tau) z^{-2\beta/3}
\tau^{2/3} z^{-\beta}\tau z^{-\beta}\dd\tau$.  The total power of $z$
should be 0 for $z$ to vanish from this expression, which gives us the
second relationship
$$
  \alpha - \frac{8}{3}\beta
=
  0
\ .
$$

Combining the two relations on $\alpha$ and $\beta$ we find
\begin{equation}
\label{eq:alpha-beta-zdep}
  \alpha = -\frac{8}{3}
\ ,\qquad\qquad
  \beta = -1
\ ,
\end{equation}
implying
\begin{equation}
\label{eq:n-selfsim-z}
  n(\sigma,z)
=
  z^{-8/3} h(\sigma/z)
\ .
\end{equation}

\subsection{Time dependent solutions}
\label{sec:self-sim:sub:z-indep}

Let us consider a self-similar distribution independent of $z$ but
dependent on time, of the form $n(\sigma,t) = \tilde t^\alpha
h(\tilde t^\beta \sigma)$, where $\tilde t = t^* -t$ and $t^*$ is a
constant, the meaning of which will become clear shortly.  The left
hand side of Eq.~\eqref{eq:kinetic} is replaced by $\partial_t n =
\alpha \tilde t^{\alpha-1} h(\tilde t^\beta \sigma) + \beta \tilde
t^{\alpha+\beta-1} \sigma h'(\tilde t^\beta \sigma)$.  Upon
introducing $\tau = \tilde t^\beta \sigma$, this becomes $\tilde
t^{\alpha-1} [\alpha h(\tau) + \beta \tau h'(\tau)]$.  The right
hand side of Eq.~\eqref{eq:subs-tau} is unchanged except for
replacing $z$ by $t$.  We thus obtain our first relationship
\begin{equation}
  \frac{7}{3}\beta - \alpha
=
  1
\ .
\label{eq:alphabeta}
\end{equation}
One could think that the second relation should come from the
conservation of mass $\int n(t,\sigma) \sigma \dd\sigma = \int
t^\alpha h(\tau) t^{-\beta}\tau t^{-\beta}\dd\tau$.  However, this
condition is incorrect because the self-similar solution in this case
gets realised only in a large\Ndash$\sigma$ tail whereas most of the
volume remains in the part which is not self-similar.  This situation
is typical of systems with finite capacity distributions, and it has
been observed previously for the Alfv\'en wave turbulence~\cite{galtier}
and for the Leith model of turbulence~\cite{cn}.  Thus, we have
$$
  n(\sigma,t)
=
  (t^*-t)^\alpha h \left(\sigma (t^*-t)^{3(\alpha +1)/7} \right)
\ .
$$
As in the case of the Alfv\'en wave turbulence~\cite{galtier}, it is
very tricky to establish how to fix the second constant $\alpha$ but
it can be found via numerical simulations of the kinetic
equation~\eqref{eq:kinetic}.

The above self-similar solution describes creation of infinitely large
particles in finite time, which rise with infinitely
large velocities. Thus, no matter how large our system is, close to
the moment $t=t^*$ there will be particles that travel across the entire
height in short time and, therefore, the $z$-independency assumption
will fail.  Note however that even close to the singularity moment
$t=t^*$ the total volume fraction of such large particles remains small.
We will study further details of such self-similar solutions 
using the ``super-local'' model in Sect.~\ref{sec:Burgers:sub:z-indep}.

\section{Locality of the self-similar solutions}
\label{sec:locality-selfsim}

%Numerical simulations show that, even if interactions are not forced
%to be local, when particles are injected in the bottom layer, a
%steady-state distribution forms that is self-similar exactly like
%in Sect.~\ref{sec:self-sim:sub:t-indep}, with the very same exponents
%as if we had imposed locality.  However the form of the solution
%appears to be quite different from that of the local case, and does
%not look like the shock-wave solution of the Burgers equation.  In
%this section we shall explore this situation.

Locality of interactions was assumed in the derivation of the
self-similar solutions in Sect.~\ref{sec:self-sim:sub:t-indep}.  This
does not need any further justification if a local collision
efficiency like in Eq.~\eqref{def:local-efcncy} is used.  However, in
the case of cut-off free interaction kernels that assumption needs to
be verified.  In order to examine its validity we will now establish
the asymptotic behavior, at small $\tau$ and at large $\tau$, of the
self-similarity function $h(\tau)$ introduced in
Sect.~\ref{sec:self-sim}.
%Since
%numerical simulations show that the self-similarity exponents $\alpha$
%and $\beta$ (cf.\ Sect.~\ref{sec:self-sim:sub:t-indep}) are the same
%as for the local interaction case,
We shall make the hypotheses (to be verified below) that at
very large $\tau$ the collision integral is dominated by contributions
of the range of much smaller $\tau$ and, conversely, that at very
small $\tau$ the collision integral is dominated by contributions of
the range of much larger $\tau$.

Let us start with the large $\tau$ case.
% which will be somewhat simpler
%as we find a power law asymptotics.  To use results of
%Sect.~\ref{sec:locality:sub:effective} we will write ... in terms of
%$\sigma$ and then pass to $\tau$ through the substitution
%Eq.~\eqref{eq:self-sim_t-indep}.  As will have to be checked once the
%solution is found, the dominant contribution to the collision integral
%comes form the non-local interaction term with small particles which for
%
Under the assumption for this range that we formulated in the previous
paragraph, the distribution in this range evolves as in
Eq.~\eqref{def:c1}, i.e.\ in the $z$-dependent steady state we have
\begin{equation}
\nonumber
  u \partial_z n
=
 -c_1 \partial_\sigma(\sigma^{4/3} n)
\ ,
\end{equation}
%(note that $c_1$ would be obtained here by taking $\sigma_{\min}$ to
%be at the scale of matching between the two asymptotic regions of our
%solution)
 which for $h(\tau)$ reduces to
\begin{equation}
\nonumber
  \tau^{2/3} [\alpha h + \beta \tau h']
=
 -c_1 \tau^{1/3} [\frac{4}{3} h + \tau h']
\ .
\end{equation}
Both sides are homogeneous in $\tau$, but the left hand side is of
degree $1/3$ higher than the right hand side, so its dominant
contribution should cancel, leading to the asymptotics $h(\tau) \sim
\tau^{-\alpha/\beta}$, and substituting values of $\alpha$ and $\beta$
from Sect.~\ref{sec:self-sim:sub:t-indep} we get $h(\tau) \sim
\tau^{-8/3}$. According to the results summarised in
Table~\ref{tab:locality-vs-scaling}, such $-8/3$ tail corresponds on
one hand to convergence of the collision integral at the large
$\sigma$ limit (as assumed in the self-similar solution) and, on the
other hand, it corresponds to dominance of interactions with much
smaller $\tau$'s as was assumed for derivations in this section.

Let us now consider the small $\tau$ range.  As we have hypothesized
above about this range, the dominant contribution to the collision
integral now comes form the non-local interaction term with large
particles, which for small $\sigma$ behaves as given in
Eq.~\eqref{def:c2}, leading to
\begin{equation}
\nonumber
  u \partial_z n
=
 -c_2 n
\ ,
\end{equation}
%(note that $c_2$ would be obtained here by taking $\sigma_{\max}$ to
%be at the scale of matching between the two asymptotic regions of our
%solution)
which for $h(\tau)$ reduces to
\begin{equation}
\nonumber
  \tau^{2/3} [\alpha h + \beta \tau h']
=
 -c_2 h
\ .
\end{equation}
This can be solved explicitly and yields
\begin{equation}
\label{eq:selfsim-tau}
  h(\tau)
=
  C_0 e^{\frac{3c_2}{2\beta} \tau^{-2/3}} \tau^{-\alpha/\beta}
=
  C_0 e^{-\frac{3c_2}{2} \tau^{-2/3}} \tau^{-8/3}
\ ,
\end{equation}
where $C_0>0$ is an integration constant and the
last member has values of $\alpha$ and $\beta$ substituted from
Sect.~\ref{sec:self-sim:sub:t-indep}.  Thanks to the very strong
stretched exponential decay of $h$ at small $\tau$ the
self-consistency of our hypotheses is straightforward to verify.  At
the same time, such fast decay at small $\tau$ ensures convergence
of the collision integral at the $\sigma=0$ limit.

We have therefore proven that our self-similar solutions are
local. Note that this result is remarkable because, in contrast with
the KZ solution, the locality property holds even without introducing
a local collisional efficiency factor.

\subsection{Numerical verification of the height dependent solutions}

\begin{figure}[ht]
\begin{center}
\includegraphics[width=.75\textwidth]{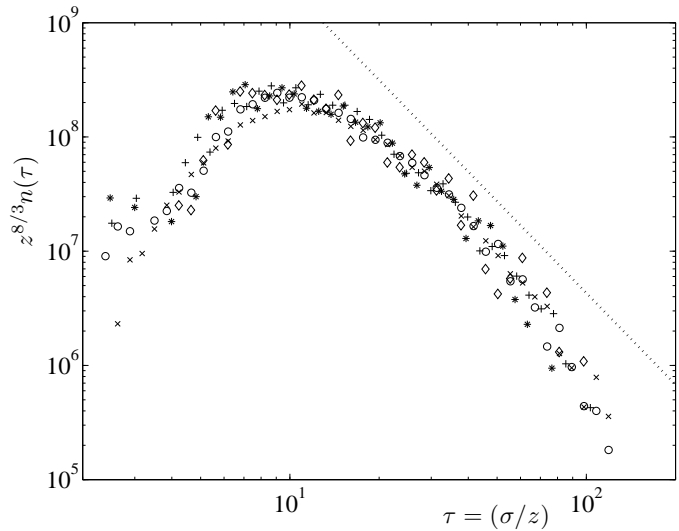} % [height=3in,width=4in]
\caption{Distribution of particle volumes after 39,000 time steps for
the situation without forced locality (``$q=\infty$''). The graph is presented
in self-similar variables according to
Eq.~\eqref{eq:n-selfsim-z}. The markers identify the spectrum for
$z=1.75$ ($\times$); $z=3.75$ ($\circ$); $z=5.75$ ($+$); $z=7.75$
($*$); $z=9.75$ ($\lozenge$). The dotted slope represents a -8/3
power law. } \label{fig:ssnonloc}
\end{center}
\end{figure}

\begin{figure}[ht]
\begin{center}
\includegraphics[width=.75\textwidth]{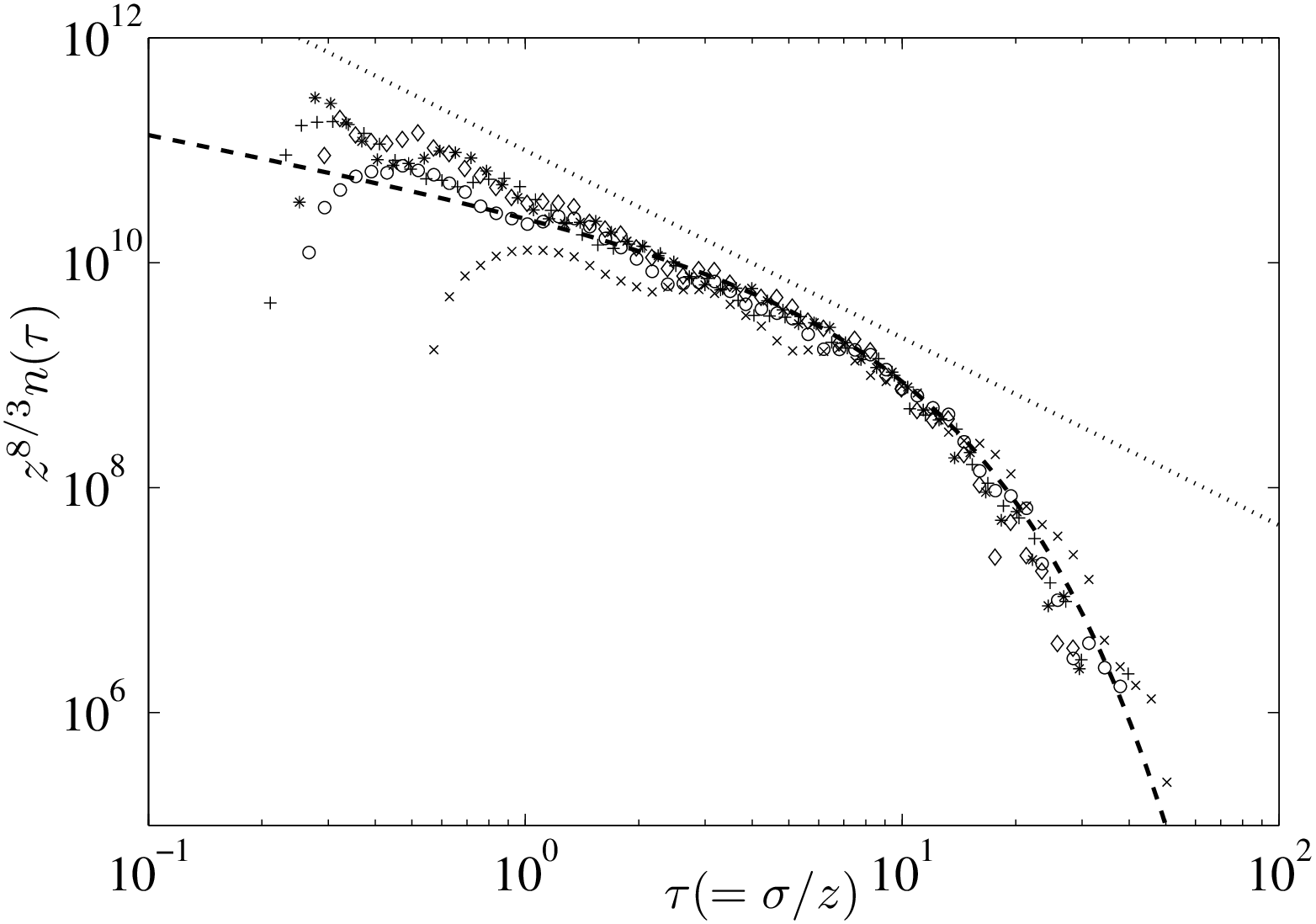} % [height=3in,width=4in]
\caption{Distribution of particle volumes after 23,000 time steps for
the forced locality situation with $q=2$. The graph is presented in
self-similar variables according to Eq.~\ref{eq:n-selfsim-z}.
The markers identify the spectrum for $z=1.75$ ($\times$); $z=3.75$
($\circ$); $z=5.75$ ($+$); $z=7.75$ ($*$); $z=9.75$ ($\lozenge$).
The dotted slope represents a $-5/3$ power law, and the dashed curve
shows $A \tau^{-2/3} \exp^{-\gamma\tau}$, 
made to fit the data at $\tau=6$.}
\label{fig:selfsimloc}
\end{center}
\end{figure}

We have performed direct numerical simulations of the set of particles
corresponding to the set-up where one should expect the self-similar
behavior. Namely, we generate particles with distribution
$n(\sigma)=\sin(\pi(\sigma-\sigma_0)/13)\sigma^{-2/3}$ and with vertical
coordinate $0<z<0.5$ and we take them out of the system as soon as 
their center has crossed the surface at $z=10$.

The results for the simulation with free merging are shown in
Fig.~\ref{fig:ssnonloc}. A rescaling to self-similar variables has already
been done. We see that profiles at different $z$ collapse, which confirms
the self-similar character of our distribution with the
self-similarity coefficients $\alpha=-8/3$ and $\beta=-1$ found in
Sect.~\ref{sec:self-sim:sub:t-indep}. Moreover, we observe that our
profile at large $\tau$ is consistent with the $-8/3$ power law found
above.

We have also performed computations with the forced locality model
as given in Eq.~\eqref{def:local-efcncy} with $q=2$. It comes to no
surprise that the observed distribution is also self-similar (since
the assumed locality has become even stronger). Naturally, the shape of
the self-similar function $h(\tau)$ is now different. It is
interesting that instead of the $-8/3$ scaling we now see a $-5/3$
slope. We will see in the next section that such a slope can be
predicted by a ``super-local'' model where the integral kinetic
equation~\eqref{eq:kinetic} is replaced by an effective differential
equation preserving the scalings of the local interactions. In the
range of large $\tau$ we observe an exponential decay $h(\tau) \sim
\exp(-b \tau) $ (where $b$ is a constant), see 
Fig.~\ref{fig:selfsimloc}.  As will be shown below, these results are also
predicted by a (regularised) ``super-local'' model.

\section{Burgers equation for local interaction case}
\label{sec:Burgers}

We will now study the systems with forced locality in greater detail
by introducing a ``super-local'' model which preserves the essential
scalings of the original kinetic equation~\eqref{eq:kinetic}, i.e.\
\begin{equation}
\label{eq:n-diffev-compl}
  \partial_t n + u \partial_z n
=
  -\sigma^{-1} \partial_\sigma (\sigma^{13/3} n^2)
\ .
\end{equation}
Particularly, Eq.~\eqref{eq:n-diffev-compl} has the same
self-similarity exponents as those found in Sect.~\ref{sec:self-sim},
in either case of height dependent or time dependent self-similar solutions.
We see that on the right hand side $n$ appears squared, making the
equation reminiscent of Burgers equation.  We are going to pursue
this idea below, by studying the simpler cases of stationary solutions
of this equation, either in $z$ or in $t$.

\subsection{Height dependent solutions}

If we look for steady state in $t$ only, then
Eq.~\eqref{eq:n-diffev-compl} reduces to
$$
  u \partial_z n
=
  -\sigma^{-1} \partial_\sigma (\sigma^{13/3} n^2)
\ .
$$
We turn this into Burgers equation by introducing new variable $s$
such that
\begin{equation}
\nonumber
  \sigma
=
  s^\lambda
\end{equation}
and the new function
\begin{equation}
\nonumber
  g(s)
=
  A s^\mu n(\sigma(s))
\ .
\end{equation}
Then $\partial_z g = -(A\lambda)^{-1} s^{\mu-8\lambda/3+1} \partial_s
(s^{13\lambda/3-2\mu} g^2)$.  If we set $\mu-8\lambda/3+1 = 0$ and
$13\lambda/3-2\mu = 0$ and $(A\lambda) = 2$ then we recover Burgers
equation:
\begin{equation}
\label{eq:Burgers_t-indep}
  \partial_z g
=
  -g \partial_s g
\ .
\end{equation}
This happens for $\lambda = 2$, $\mu = 13/3$ and $A = 1$.

Conservation of total particle volume leads to the conservation of the integral
$\int g(s) \dd s$, and we deal with the usual Burgers dynamics 
even for the weak solutions (i.e.\ any regularisation of this equation should
conserve the volume).  In this case we get no finite-time singularity
since $A$ and $\lambda$ are positive.  We will use the analogy of \eqref{eq:Burgers_t-indep} with Burgers equation and assume a discontinuity in our function $g$ would be a shock in the equivalent Burgers system. 
The sawtooth shock can be seen
to evolve such that at ``time'' $z$ the shock is at $s_* \sim z^{1/2}$
and its height is $g_* \sim z^{-1/2}$ (hint: write $\dd s_* / \dd z =
g_*/2$ and $s_* g_* = B$ where $B$ is constant). For the original
variables this gives $\sigma_* \sim z^{\lambda/2} = z$ and $n_0 \sim
z^{-\mu/2} z^{-1/2} = z^{-8/3}$.  One then sees that this solution is
self-similar with the scaling we have found above.  In fact
\begin{equation}
\nonumber
  n(\sigma,z)
=
  \begin{cases}
    z^{-8/3} (\sigma/z)^{-5/3} & \text{if $\sigma \leq z$,} \\
    0 & \text{if $\sigma > z$.}
  \end{cases}
\end{equation}
%Note that the $\sigma^{-5/3}$ scaling seems to give rise to non-local
%interactions with large particles so this approach is only valid if we
%impose locality of interaction.

Remarkably, the $-5/3$ scaling of the self-similar function
$h(\tau)$ is indeed observed in the numerical simulation of the
particles with the forced locality collision efficiency, see 
Fig.~\ref{fig:selfsimloc}. This fact indicates that, in spite of
simplicity, the super-local model~\eqref{eq:n-diffev-compl} is
indeed quite efficient in predicting certain essential features of
the particle kinetics. However, we have not observed any signature of
a shock in our numerical results. Such a shock should be considered
as an artifact of super-locality which is smeared out when a finite
interaction range is allowed.

In fact, following the method exposed in Sect.~4.2 of
ref.~\cite{dong2}, it is also possible to obtain the asymptotic
behaviour of $n(\sigma,z)$ for large $\tau = \sigma/z$ (see
Sect.~\ref{sec:self-sim:sub:t-indep}).  This is beyond the reach of
the Burgers model~\footnote{Even if we added diffusive regularization to
the Burgers model to account for not strict super-locality, we would
get the incorrect $z^{-8/3} \exp(-\gamma\sigma/z)$ behaviour, where
$\gamma>0$ is some constant (see also Appendix~\ref{app:flux}).}.
%% This suggests that the model~\eqref{eq:n-diffev-compl} could be
%% further improved by adding a diffusion regularisation to the Burgers
%% equation, so that the total volume would still be preserved, but the
%% shock front would be smeared to relax the degree of locality.  For the
%% case of viscous Burgers equation one could use the Hopf-Cole transform
%% to solve exactly the evolution equation and compute the asymptotics of
%% the large $s$ tail, and see that it is dominated by the diffusive
%% term.  This should be the case more generally and thus we get that the
%% large $s$ tail of $g(s,z)$ behaves like $z^{-1/2} \exp(-\gamma
%% s^2/z)$, or in the original variables, the large $\sigma$ tail of
%% $n(\sigma,z)$ behaves as $z^{-8/3} \exp(-\gamma\sigma/z)$, where
%% $\gamma$ is some positive constant. In fact this result seems to be
%% quite robust and valid not only for the diffusive generalisation of
%% the Burgers equation but also for other kind of regularisations.  The
%% exponential shape of the self-similar function $h(\tau) \sim
%% \exp(-\gamma\tau)$ is indeed in a very good agreement with our
%% numerical results for the particle kinetics with forced locality, see
%% Fig.~\ref{fig:selfsimloc}.
Following ref.~\cite{dong2} and using notation from our
Sect.~\ref{sec:self-sim:sub:t-indep}, we introduce the ansatz $h(\tau)
\sim A \tau^{-\theta} e^{-\gamma\tau}$, where $A$, $\gamma$ and
$\theta$ are real constants, of which we shall only determine $\theta$
here.  With this ansatz and using the flux formulation described in
Appendix~\ref{app:flux}, in particular Eqs.~\eqref{eq:flux} and
\eqref{eq:dt-div-flux}, we can write Eq.~\eqref{eq:subs-tau} as (note
that we take the values of $\alpha$ and $\beta$ from
Eq.~\eqref{eq:alpha-beta-zdep}):
\begin{multline*}
  \tau^{2/3}
  [\textstyle{-\frac{8}{3}} A \tau^{-\theta} e^{-\gamma\tau}
       +(\theta-\gamma\tau) A \tau^{-\theta} e^{-\gamma\tau}]
=\\
  \tau^{-1} \partial_\tau
  \int_0^\tau \!\dd\tau_1
    \int_{\tau-\tau_1}^\infty \hspace{-2ex}\dd\tau_2 \ 
      K(\tau_1,\tau_2)
      A^2 \tau_1^{1-\theta} \tau_2^{-\theta}
      e^{-\gamma(\tau_1+\tau_2)}
\end{multline*}
The left hand side scales as $\tau^{2/3-\theta} e^{-\gamma\tau}$ while
the right hand side can be seen to scale, for large $\tau$, as
$\tau^{4/3-2\theta} e^{-\gamma\tau}$ (in order to see this, note that
$e^{-\gamma(\tau_1+\tau_2)}$ attains its maximum over the integration
domain along the segment $\tau_1+\tau_2=\tau$ with $\tau_1,\tau_2 >0$
and becomes much smaller for $\tau_1+\tau_2-\tau \gtrsim \gamma^{-1}$,
so that the {\em effective} integration domain is a band of width of
order $\gamma^{-1}$ around the segment $\tau_1+\tau_2=\tau$).  In
order for the two sides to have the same scaling we must have $\theta
= 2/3$.  Then $h(\tau) \sim A \tau^{-2/3} e^{-\gamma\tau}$ and
$n(\sigma,z) \sim A z^{-2} \sigma^{-2/3} e^{-\gamma\sigma/z}$.

\subsection{Time dependent solutions}
\label{sec:Burgers:sub:z-indep}

Let us now seek $z$\textendash independent solutions of
Eq.~\eqref{eq:n-diffev-compl}.  In this situation the latter reduces
to
$$
  \partial_t n
=
  -\sigma^{-1} \partial_\sigma (\sigma^{13/3} n^2)
\ .
$$
We turn this into Burgers equation as above, introducing $s$ and
$g(s)$ as above.  Then $\partial_t g = -(A\lambda)^{-1}
s^{\mu-2\lambda+1} \partial_s (s^{13\lambda/3-2\mu} g^2)$.  If we set
$\mu-2\lambda+1 = 0$ and $13\lambda/3-2\mu = 0$ and $A\lambda = 2$
then we recover Burgers equation.  This happens for $\lambda = -6$,
$\mu = -13$ and $A = -1/3$.

In order to know what happens at shocks we need to know what
quantity is conserved by evolution, even at shocks.  We know that the
original system conserves the volume $\int n \sigma \dd\sigma$, which
translates for $g$ to conservation of $(\lambda/A) \int g(s)
s^{2\lambda-\mu-1} \dd s$, and since $2\lambda-\mu-1 = 0$ this
simply means conservation of $\int g(s) \dd s$.  Thus once again we
really deal with the usual Burgers dynamics.

If the initial distribution of $n$ is peaked around $\sigma_0$ with
height $n_0$ then the initial distribution of $g$ is peaked around
$s_0 = \sigma_0^{1/\lambda}$ with height $g_0 = A s_0^\mu n_0$.  It is
convenient to suppose that the peak is of compact support, say between
$\sigma_1 < \sigma_2$, corresponding to $s_1 > s_2$.  Since $n$ (the
particle density) is positive but $A$ is negative, $g$ will be negative
and shocks will move towards smaller $s$.  The peak evolves to give a shock,
which will have formed at some $s>s_2$.  To good approximation we get
a single sawtooth shock which moves towards 0 and reaches it in finite
time, which for $n$ means (since $\lambda < 0$) that there is a
finite-time singularity at infinite volume.

The important feature is that the shock in $g$ will arrive at $s=0$
at some finite time $t^*$, and for $t$ close to $t^*$ its height and
speed are approximately constant, say height $g^*$ and position
$s=\tilde t w^*$ where $\tilde t = t^*-t$.  This translates for $n$
to a jump of height $A^{-1} s^{-\mu} g^* = A^{-1} (\tilde t
w^*)^{-\mu} g^* \propto \tilde t^{-\mu}$ at position $\sigma =
s^\lambda \propto \tilde t^\lambda$.  This is compatible with
self-similarity $n(\sigma,t) = \tilde t^\alpha h(\tilde t^\beta
\sigma)$ only for exponents $\alpha=-\mu=13$ and $\beta=-\lambda=6$,
which satisfy the condition from Eq.~\eqref{eq:alphabeta}.

Note also that, since $g$ can be considered to be approximately
constant behind the shock (i.e.\ towards large $s$) , the distribution of
$n$ behind the jump (i.e.\ towards small $\sigma$) is like $\sigma^{-13/6}$,
which is a finite capacity power law, as required by conservation of
total initial finite mass.

Since self-similarity only appears in the tail of the distribution,
and the tail has finite capacity, it is difficult to obtain good
statistics in numerical simulations for this model. In the tail,
there will be very large particles, but the void fraction will be
large too, as $\int n \sigma d\sigma$ is constant, resulting in a
sparse data set in the numerical simulation.

\section{Concluding remarks}

As we have seen, the very simple model in which particles move at
their terminal velocity and merge upon collision appears to be very
rich in features.  For this model, we have derived the
Smoluchowski kinetic equation~\eqref{eq:kinetic} 
with a kernel for differential sedimentation.

First of all, we considered a setup analogous to one used in turbulence theory
where small particles are produced and large particles are removed from the 
system with a wide inertial interval in between these source and sink scales.
We obtained a KZ spectrum (Fig.~\ref{fig:kolzakh}) and showed that it
is relevant for the systems with forced locality but irrelevant in the
free-merging case.  In the latter case we derived a
model~\eqref{eq:NL-effkin} in which the dominant interactions are
non-local and we obtained its steady state solution in
Eq.~\eqref{eq:n-NLNL}, which was verified with DNS
(Fig.~\ref{fig:nonloc}).

We have also considered self-similar solutions which are either
height dependent or time dependent.  This was
done for both the kinetic equation~\eqref{eq:kinetic} and for a
model with ``super-local'' interactions~\eqref{eq:n-diffev-compl}.
For the time dependent dynamics, we predicted a finite-time creation of
infinitely large particles. The solutions for height dependent dynamics
were verified with DNS. Although most particle distributions in the 
atmosphere are height dependent~\cite{pruppacher}, the relevance
of self-similarity in such distributions requires further study.

Our theoretical results were obtained from the kinetic
equation~\eqref{eq:kinetic} which is essentially a mean field
approach.  Thus, it is intriguing that such theoretical predictions in
all considered situations agree well with the numerical simulations of
the complete system.  This suggests that the mean field assumption
leading to the kinetic equation should be valid in the considered
sedimentation model, and the origin of this could be addressed in the
future with techniques of field theory and renormalization.

Finally, we have only considered very simple models either without the
collision efficiency factor, or with a simple forced locality factor
conform Eq.~\eqref{def:local-efcncy}.  Other forms of localizing
kernels should be considered for more realistic situations.

\subsection*{Acknowledgements}
\label{sec:acknowledge}

We would like to thank Miguel Bustamante, Antti Kupiainen, Lian-Ping Wang and Oleg
Zaboronski for helpful discussions and suggestions.

\appendix

\section{Locality of power-law distributions}
\label{sec:locality}

Power law distributions of the form $n(\sigma) \sim \sigma^\nu$ are
important because they arise from the formal analysis of the KZ
spectra, self-similar solutions, etc.  However, some of such formal
considerations implicitly use convergence of the collision integral on
RHS of Eq.~\eqref{eq:kinetic} which has the meaning of the interaction
{\em locality}.  Conversely, other derivations may assume {\em
non-locality} i.e.\ that the evolution is dominated mostly by the
interactions with the smallest or the largest particles in the system
corresponding to the vicinities of the small-$\sigma$ and the
large-$\sigma$ integration limits.  Therefore, the conditions of
convergence of the collision integral must be found, and this will be
done in this appendix for a general distribution $n(\sigma) \sim
\sigma^\nu$.

%For most of our considerations below, it will be important whether in
%the collisions integral appearing on the RHS of
%Eq.~\eqref{eq:kinetic}, for a given $\sigma$ the dominant contribution
%to the integral comes from $\sigma_1$, $\sigma_2$ of the order of
%$\sigma$ -- the local case, -- or on the contrary from values of the
%integration variables close to the lower bound of integration -- lower
%non-local case, -- or upper bound of integration -- the upper non-local
%case.  In this Section we develop such considerations for the case
%when the density $n$ is supposed to have a scaling behaviour of the
%form $n(\sigma) \sim \sigma^\nu$.  We immediately point out that in
%several situations of interest to us, treated in the text, $n$ does
%not follow such a simple scaling, and further analysis is needed then.

%\label{sec:locality:sub:scaling}

%We first determine if the interaction described by the collision
%integral on the RHS of Eq.~\eqref{eq:kinetic} is local or non-local,
%as a function of the scaling exponent $\nu$ of particle density
%$n(\sigma) \sim \sigma^\nu$.

Introduce $f(\sigma_1,\sigma_2) = K(\sigma_1,\sigma_2) n_1 n_2$.  
Equation~\eqref{eq:kinetic} may be expressed in terms of
$f$ as
$$
  \frac{\dd}{\dd t} n
=
  \int_{\sigma_{\min}}^{\sigma/2} \dd\sigma_1
    f(\sigma_1,\sigma-\sigma_1)
 -\int_{\sigma_{\min}}^{\sigma_{\max}} \dd\sigma_1
    f(\sigma_1,\sigma)
\ .
$$
We can then split $\dd n/\dd t$ into two parts, which we shall call
lower and upper contributions:
$$
  \frac{\dd}{\dd t} n
=
  \left. \frac{\dd}{\dd t} \right|_< n
 +\left. \frac{\dd}{\dd t} \right|_> n
$$
with
\begin{align*}
  \left. \frac{\dd}{\dd t} \right|_< n
&=
  \int_{\sigma_{\min}}^{\sigma/2} \dd\sigma_1
    [f(\sigma_1,\sigma-\sigma_1)
    -f(\sigma_1,\sigma)]
\\
  \left. \frac{\dd}{\dd t} \right|_> n
&=
 -\int_{\sigma/2}^{\sigma_{\max}} \dd\sigma_1
    f(\sigma_1,\sigma)
\ .
\end{align*}

We start by analyzing the lower contribution, more specifically its
convergence as $\sigma_{\min}$ goes to 0.  For this the value of the
integrand at $\sigma_1 \ll \sigma$ needs to be known.  This can
be approximated by the Taylor expansion
$$
  f(\sigma_1,\sigma-\sigma_1)
 -f(\sigma_1,\sigma)
\sim
  \sigma_1 \partial_\sigma f(\sigma_1,\sigma)
\ .
$$
%
%% The contribution to the collision integral from the particles having
%% small size, is well approximated by
%% $$
%%   \left. \frac{\dd}{\dd t} \right|_\ll n
%% =
%%   \int_{\sigma_{\min}}^{\sigma/2} \dd\sigma_1
%%     [f(\sigma_1,\sigma-\sigma_1)
%%     -f(\sigma_1,\sigma)]
%% \approx
%%   \int_{\sigma_{\min}}^{\sigma/2} \dd\sigma_1
%%     \sigma_1 \partial_\sigma f(\sigma_1,\sigma)
%% \ .
%% $$
%
%For the case of interest of $\sigma_1$ small, we can Taylor expand
%$f$ and have $f(\sigma_1,\sigma-\sigma_1) - f(\sigma_1,\sigma) \sim -
%\sigma_1 \partial_2 f(\sigma_1,\sigma)$.
For small $\sigma_1$ we also have $f(\sigma_1,\sigma) \sim c \kappa^4 \pi
\sigma^{4/3} n_1 n$ so we have
%$f(\sigma_1,\sigma-\sigma_1) - f(\sigma_1,\sigma) \sim - n_1 \sigma_1
%\partial_\sigma (\sigma^{4/3} n)$.  Finally
$$
  \left. \frac{\dd}{\dd t} \right|_< n
\approx
 - c \kappa^4 \pi
  \left[\int_{\sigma_{\min}}^{\sigma/2} n_1 \sigma_1 \dd\sigma_1 \right]
  \partial_\sigma (\sigma^{4/3} n)
\ .
$$
The interaction is local at small scales iff the integral above
remains finite when $\sigma_{\min} \to 0$.  This is equivalent to $\nu
> -2$.

We now turn to the upper contribution, more specifically its
convergence as $\sigma_{\max}$ goes to infinity.  For this the value
of the integrand at $\sigma_1 \gg \sigma$ needs to be known.  In
these asymptotics we have $f(\sigma_1,\sigma) \sim \sigma_1^{4/3} n_1
n$ and therefore
$$
  \left. \frac{\dd}{\dd t} \right|_> n
\approx
 -n \int_{\sigma/2}^{\sigma_{\max}} n_1 \sigma_1^{4/3} \dd\sigma_1
\ .
$$
The interaction is local at large scales iff the integral above
remains finite when $\sigma_{\max} \to \infty$.  This is equivalent to
$\nu < -7/3$.

We thus get the picture that for $\nu < -7/3$ the interaction is local
at large scales but non-local at small scales.  For $-7/3 \leq \nu
\leq -2$ both ends are non-local.  And for $\nu > -2$ interaction is
non-local at large scales but local at small scales.  In particular, we
never have locality at both ends.
\bigskip

\begin{table}
\centerline{
\begin{tabular}{|l|c|c|c|}
\hline
  &&&\\[-1.5ex]
  & $\nu < -\frac{7}{3}$ & $-\frac{7}{3} \leq \nu \leq -2$ & $-2<\nu$
  \\[.7ex]
\hline
  upper & local & \multicolumn{2}{|c|}{non-local} \\
\hline
  lower & \multicolumn{2}{|c|}{non-local} & local \\
\hline
%%   upper & local & non-local & non-local \\
%%   lower & non-local & non-local & local
\end{tabular}
}
\caption{Locality of interaction with small and large particles, as
  dependent on the scaling exponent of $n(\sigma)$ (compare Connaughton et al.~\cite{cnzab}).}
\label{tab:locality-vs-scaling}
\end{table}

\section{Considerations on the flux}
\label{app:flux}

The flux $\Phi(\sigma)$ of volume going into particles of volume larger than
$\sigma$ can be obtained by the following consideration.  The flux in
question is the volume contained in particles of volumes smaller $\sigma$ that
merge during unit time with some particle to give a particle of volume larger
than $\sigma$.  Say one such particle has $\sigma_1<\sigma$, then it
can merge with any particle with $\sigma_2$ such that $\sigma_1 + \sigma_2
> \sigma$, i.e.\ $\sigma_2 > \sigma-\sigma_1$.  Using the collision kernel
$K$ the above consideration is made formal as
\begin{equation}
\label{eq:flux}
  \Phi(\sigma)
=
  \int_0^\sigma \dd\sigma_1 \int_{\sigma-\sigma_1}^{\infty} \dd\sigma_2\,
    \sigma_1 K(\sigma_1,\sigma_2) n(\sigma_1) n(\sigma_2)
\ .
\end{equation}
One readily verifies by direct computation (and a minor trick) that
the right hand side of the kinetic equation~\eqref{eq:kinetic} equals
$-\sigma^{-1} \partial_\sigma \Phi(\sigma)$, so we have as we may expect
\begin{equation}
\label{eq:dt-div-flux}
  \sigma \frac{\dd n(\sigma)}{\dd t}
=
 -\partial_\sigma \Phi(\sigma)
\ .
\end{equation}
We immediately remark two things about $\Phi$.  First, it is convergent at
the lower bound ($\sigma_1 \to 0$) if and only if interaction with the small $\sigma$
tail is local, and similarly it is convergent at the upper bound ($\sigma_2
\to \infty$) if and only if interaction with the large $\sigma$ tail is local
(compare with Appendix~\ref{sec:locality}).

The other remark is that $\Phi(\sigma)$ scales as
$\sigma^{4/3+3+2\nu}$ (if $n$ scales as $\sigma^\nu$).  Hence, for $\nu
= -(4/3+3)/2 = -13/6$ we have $\partial_\sigma \Phi(\sigma) = 0$ and
thus, from Eq.~\eqref{eq:dt-div-flux}, $\sigma^{-13/6}$ is a
stationary power law solution.

The next thing we do is Taylor expand $n$ around $n(\sigma)$ in the
expression~\eqref{eq:flux} of the flux.  Then to lowest (zeroth) order we
get
\begin{equation*}
  \Phi(\sigma)
=
  n(\sigma)^2
  \int_0^\sigma \dd\sigma_1 \int_{\sigma-\sigma_1}^{\infty} \dd\sigma_2\,
    \sigma_1 K(\sigma_1,\sigma_2)
\ .
\end{equation*}
Since the integral above scales as $\sigma^{4/3+3}$, this can be
written as 
\begin{equation*}
	\Phi(\sigma) 
= 
	C^\prime \sigma^{4/3+3} n(\sigma)^2
\ ,
\end{equation*}
with
$C^\prime>0$ (since $K\geq0$), and substituting this into
Eq.~\eqref{eq:dt-div-flux} we get an equation equivalent to Burgers
equation (cf.\ Sect.~\ref{sec:Burgers}).

Remark: perhaps one caveat is that the simple Taylor expansion
proposed above doesn't seem to correspond to an expansion in some
small parameter of the problem.  One natural small parameter could be
$q-1$ from the definition~\eqref{def:local-efcncy} of the collision
efficiency.  But the expansion in $q-1$ would be slightly more
complex.

One can carry on this Taylor expansion and get terms of higher order,
which will have more derivatives $\partial_\sigma$ and higher powers
of $\sigma$.  In the ``Burgers'' coordinates introduced in
Sect.~\ref{sec:Burgers}, the same holds but with powers and
derivatives in $s$.  In particular to next order we get, in the setup
of Sect.~\ref{sec:Burgers:sub:z-indep}, $\partial_t g =
-\partial_s(C_1 g^2 + C_2 s\partial_s g^2)$.

\bibliographystyle{nature}
\bibliography{paperV}

\end{document}